# Inverse Solidification Induced by Active Janus Particles

Tao Huang[1,2], Vyacheslav R. Misko[3,4], Sophie Gobeil[1], Xu Wang[5], Franco Nori[3,6], Julian Schütt[1,5], Jürgen Fassbender[5], Gianaurelio Cuniberti[1], Denys Makarov[5], Larysa Baraban[1,2]

[1]*Max Bergmann Center of Biomaterials and Institute for Materials Science, Technische Universität Dresden, 01062 Dresden, Germany*
[2]*Helmholtz-Zentrum Dresden-Rossendorf e.V., Institute of Radiopharmaceutical Cancer Research, Bautzner Landstrasse 400, 01328 Dresden, Germany*
[3]*Theoretical Quantum Physics Laboratory, RIKEN Cluster for Pioneering Research, Wako-shi, Saitama 351-0198, Japan*
[4]*µFlow group, Department of Chemical Engineering, Vrije Universiteit Brussel, Pleinlaan 2, 1050 Brussels, Belgium*
[5]*Helmholtz-Zentrum Dresden-Rossendorf e.V., Institute of Ion Beam Physics and Materials Research, Bautzner Landstrasse 400, 01328 Dresden, Germany*
[6]*Physics Department, University of Michigan, Ann Arbor, Michigan 48109-1040, USA*

**Abstract.** Crystals melt when thermal excitations or the concentration of defects in the lattice is sufficiently high. Upon melting, the crystalline long-range order vanishes, turning the solid to a fluid. In contrast to this classical scenario of solid melting, here we demonstrate a counter-intuitive behavior of the occurrence of crystalline long-range order in an initially disordered matrix. This unusual solidification is demonstrated in a system of passive colloidal particles accommodating chemically active defects – photocatalytic Janus particles. The observed crystallization occurs when the amount of active-defect-induced fluctuations (which is the measure of the effective temperature) reaches critical value. The driving mechanism behind this unusual behavior is purely *internal* and resembles a blast-induced solidification. Here the role of "internal micro-blasts" is played by the photochemical activity of defects residing in the colloidal matrix. The defect-induced solidification occurs under non-equilibrium conditions: the resulting solid exists as long as a constant supply of energy in the form of ion flow is provided by the catalytic photochemical reaction at the surface of active Janus particle defects. Our findings could be useful for understanding of the phase transitions of matter under extreme conditions far from thermodynamic equilibrium.

## 1. Introduction

Macroscopic properties of solids are determined by their crystalline structure and its purity. Introducing defects in the lattice, by doping or increasing the temperature, results in the displacement of atoms of the initial crystal from their equilibrium positions. When these displacements exceed some critical value, the crystal melts turning to a liquid (Figure 1a).[1] Typically, defect-induced solid melting occurs via defect proliferation along the crystal-lattice directions, leading to the increase of the density of defects.[2]

Going beyond this classical picture, by driving a solid out of equilibrium, enables exciting fundamental discoveries, including the observation of solidification in an externally-driven system, called "freezing-by-heating" transition,[3] or re-entrant amorphous-to-solid transitions under extreme pressure generated by a blast.[4] Although these findings primarily concern materials science, similar non-equilibrium processes occur in various disciplines, from behavioral zoology to traffic research.[5] In the case of condensed matter systems, non-equilibrium processes are important for the realization of materials with unique physical and chemical properties in terms of density and phase formation, which cannot be achieved using conventional fabrication. To explore in full the potential of non-equilibrium solidification for material science, the matter should be exposed to extreme conditions such as ultra-high pressure or extremely high and fast changing environmental temperature. This renders the study of the crystallization or melting processes out of equilibrium to be a demanding yet very challenging task, which requires specific experimental settings and advanced infrastructure. Therefore, it is insightful to find a proper model system, which will allow mimicking experiments of matter under extreme conditions yet in a standard laboratory environment.

Soft matter systems of colloidal particles have been successfully studied for the macroscopic modeling of condensed matter phenomena, including melting-to-freezing transitions,[6] formation of crystals,[7] clusters,[8] and colloidal glasses, as well as the Kosterlitz-Thouless transition specific to two dimensional (2D) materials.[1c, 7c] It is established that the melting of colloidal crystals is driven by the propagation of stress fronts along the symmetry axes of crystalline monolayers.[9] The observed behavior in a soft matter system is similar to the defect-induced melting processes occurring in solids.[2] Using active colloidal systems, like Janus particles, rod-like or tubular micromotors[10] instead of passive colloidal beads, allows modeling a broad range of non-equilibrium phenomena ranging from biological self-organization and cooperative behavior[11] to crystallization and diffusion in condensed matter.[9, 12] These studies explore the interactions between active particles,[13] active and passive beads,[12, 14] and active particles and walls.[10d, 15] In particular, major progress has been achieved in the understanding of a collective behavior,[16] phase transitions in a dense suspension of self-propelled colloids,[17] dynamic pattern formation, and active turbulence.[18] Although soft matter platforms are broadly used to study emergent material science problems, the modeling of inverse phase transitions in active matter out of equilibrium have not been addressed yet.

Here, we demonstrate – in experiments and in numerical simulations – a striking counter-intuitive behavior: the occurrence of long-range order in an initially disordered suspension of colloidal particles (Figure 1b and c). The observed ordering is induced by active defects, namely catalytic Janus particles, dispersed in a passive matrix of colloidal beads. Crystallization in the colloidal suspension is observed when the concentration of active defects increases and reaches a critical value. This situation is opposite to the classical defect-induced crystal melting observed in condensed[2] and soft matter systems.[19] This allows us to refer to the observed active-defect-induced solidification as to an *inverse liquid-to-solid transition* under increasing fluctuations.

To understand the mechanism responsible for the observed unusual behavior, we emphasize its similarity to the amorphous-to-crystalline phase transition in solids under extremely high pressure, produced by a compression wave generated by a blast. Similar to the shock waves generated by explosions or gas-propelled projectiles used to produce ultra-high transient pressure (e.g., for diamond synthesis,[20] biomolecular condensates,[21] or studies of emergent phase transitions[13]), active defects in our experiments generate compression waves by ejecting products of the catalytic photochemical reaction from the surface of Janus particles when illuminated by light (Figure 1c). This leads to the compression of the surrounding colloidal matrix driven by the stimuli, which are internal to the system. When the concentration of active defects is sufficiently high – such that the compressed regions, generated by neighboring defects overlap – then the colloidal matrix undergoes the induced solidification transition.

This unusual solidification due to *internal* drive has not been previously reported. Although we note a similarity of our observation to the freezing-by-heating transition (both processes imply far from equilibrium conditions and such ingredients as the existence of driving, repulsive interaction among the species and confinement), the essential feature of the transition reported here is that it is induced by internal active defects and not by an external drive (cf., e.g., Refs.[7d, 8d] where colloidal aggregation was induced by external fluid flows).

Thus, we propose a novel route towards crystallization of matter under conditions far from equilibrium. The reported system involving active defects is a versatile playground to address the behavior of materials under extreme conditions, yet available in standard laboratories using convenient soft matter models.

## 2. Results

We investigate the interaction of visible-light-driven Ag/AgCl-based Janus particles[22] (diameter 2 μm), with a dense environment of passive $SiO_2$ beads (diameter 2 μm) in pure water. Janus particles act as active defects introduced into a dense amorphous colloidal matrix (Figure 1c). Note that, depending on the density of the matrix, the colloids in an initial state can show properties of either a liquid or an amorphous solid (when the Brownian motion of colloidal particles is mainly restricted by the closest neighbors). Since we focus on the transition from a state possessing no crystalline order to a crystal solid, we will often refer to the initial state as being "amorphous", although this also applies to a liquid.

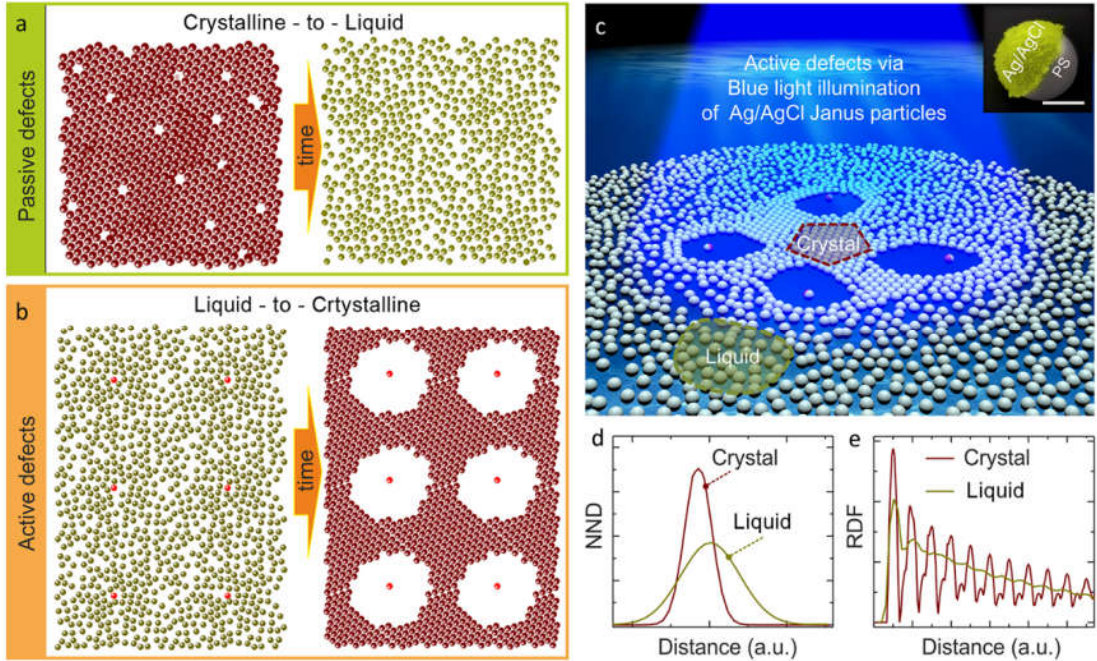

**Figure 1.** Concept of the liquid-to-crystalline phase transition due to internal drive. (a) Sketch illustrating a conventional defect-induced solid melting: a transition from crystalline to liquid state. (b) Liquid-to-crystalline transition induced by active defects. (c) Schematics of the experimental realization of the liquid-to-crystalline transition using a soft matter colloidal system. The crystallization is a result of the critical compression of the initially disordered colloidal matrix due to the internal pressure stemming from the photochemical activity of blue light illuminated Ag/AgCl Janus particles (shown with the red caps). Inset in c: scanning electron microscopy (SEM) image of polystyrene (PS) based Janus particle where the Ag/AgCl cap is shown with false color. Scale bar, 1 μm. (d) Nearest neighbor distribution (NND) function for liquid and crystalline states. The NND peak becomes sharper and shifts to smaller distances when the system is compressed and undergoes the transition from liquid to crystalline state. (e) Radial distribution function (RDF) showing one peak followed by a featureless tail if the system is in the liquid state. RDF with many sharp peaks is typical for a crystalline state.

In contrast to the previous works,[12-14] we rely on catalytically active but *immobile* Janus particles, which are fixed to the surface. Although not moving, they do release charged species into the water upon the photocatalytic reaction of decomposition of AgCl under blue light illumination[18a, 22-23]. We explore the exclusion phenomenon between active defects and passive beads that leads to the compression of the passive matrix and can result in a phase transition of the initially amorphous (or liquid) colloidal system, showing only short-range order, into a crystalline state characterized by long-range order. Since the location and orientation of each active defect is fixed, we focus on the dynamics of the surrounding passive particles that reflect transient processes in the system. To quantitatively access the relevant processes, we follow the spatial distributions of passive particles in the system and analyze the mean squared displacement (MSD), nearest neighbors distribution (NND) and radial distribution function (RDF). Figures 1d and 1e show the RDF and NND of amorphous and crystalline systems.

**Individual active defects:** Figures 2a to 2h summarize the behavior and transient regimes in the system containing active and passive particles. When illuminated with a reference white light (that does not cause a photocatalytic reaction), passive beads surrounding the Janus particle reveal Brownian diffusion and interact via repulsive electrostatic interaction due to their negative zeta potentials ($\zeta_{JP}$ = - 15.1 ± 3.6 mV, $\zeta_{PP}$ = - 38.7 ± 0.9 mV). The coverage of the surface with passive beads remains homogeneous with a nominal areal density of 0.12 μm$^{-2}$.

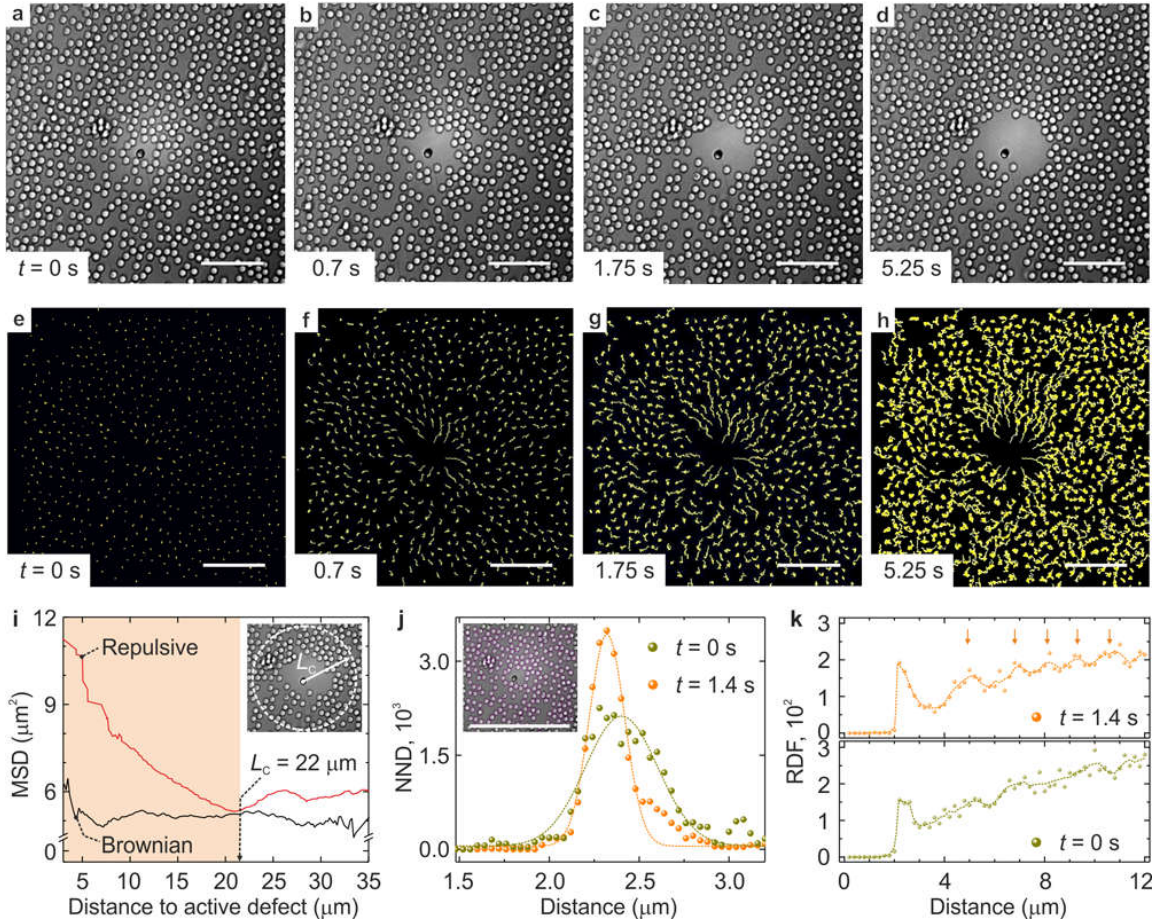

**Figure 2.** Compression of a colloidal matrix of passive beads by active defects. (a-d) A sequence of optical microscopy images showing the time evolution of the system of immobile, yet photocatalytically active, Janus particles interacting with the surrounding dense passive matrix (SiO$_2$ beads) under blue light illumination. The repulsive interaction between the Janus particle and beads induces a local compression wave in the passive matrix. Scale bar, 20 μm. (e-h) The time evolution of the recorded trajectories of passive beads showing a local compression of the passive matrix: the most substantial repulsion is observed between the close-neighbor passive beads and the Janus particle. Scale bar, 20 μm. (i) The mean squared displacement (MSD) averaged over 500 passive beads as a function of the distance to the active defect. The local nature of the impact of the active defect on the surrounding matrix is illustrated by comparing the MSD due to defect-induced motion with the MSD due to thermal Brownian motion in 8 s. Passive beads located at a distance larger than $L_c$ = 22 μm from the active defect are not affected by the presence of the catalytically active defect. (j) The NND analysis of the change in the density of the passive matrix in the neighborhood of the active defect at $t$ = 0 and $t$ = 1.4 s. As a result of the local compression, the NND peak sharpens and shifts toward smaller distances. (k) The RDF for the initial amorphous state (time $t$ = 0 s) and for the locally compressed state of the passive matrix ($t$ = 1.4 s). The compression results in the onset of a long-range order in the system: the RDF acquires a number of well-resolved peaks shown by orange arrows in the upper panel of (k). Still, the system remains in the amorphous state.

Once the blue light illumination is turned on, the photocatalytic decomposition of Ag/AgCl at the cap of Janus particles sets in[22] and results in a local release of charged species in the solution. This process generates an electric field around the active particle pointing towards the cap. The field acts electrophoretically on the negatively charged passive silica beads and generates their repulsion from the active Janus defects.[22] This leads to a *compression* of the amorphous colloidal matrix that is associated with the local decrease of the interparticle distances around the Janus particle. The induced mechanical stress is absorbed by the dense passive colloidal matrix and propagates radially from the Janus defect as shown in Figure 2a-d. Furthermore, we can identify the resulting ring-shaped area around the active defect, where the density of passive beads is higher than the nominal density of 0.12 $\mu m^{-2}$. By analyzing the MSD data taken of 500 passive beads under white and blue light illumination (black and red curves in Figure 2i, respectively), we determine that the displacement of passive beads, located at distances larger than 22 µm from the active defect, does not exceed the Brownian diffusion. Therefore, these particles are not affected by the presence of the catalytically active defect. To illustrate this, we draw a circle of radius $L_c$ = 22 µm around a Janus particle (inset in Figure 2i, frame is taken at $t$ = 5.25 s) indicating the area where the local density of passive beads is not homogenous. Within this circle, there is a ring-shaped region where the density of passive beads is increased up to 0.13 $\mu m^{-2}$. The density of passive beads in the vicinity of the active defect is close to zero. In the following, the parameter $L_c$ is used to determine the critical distance between active Janus particles and to investigate the compression of the colloidal matrix.

We quantify the displacement of passive beads at different observation times, $t$, after turning the blue light on, using NND (Figure 2j) and RDF (Figure 2k). This analysis is carried out in the area of interest within 22 µm from the active defect. Figure 2j demonstrates the NND calculated at $t$ = 0 s (green curves) and $t$ = 1.4 s (orange curve). For blue light illumination at $t$ = 1.4 s, the NND peak becomes sharper and shifts towards the smaller distance. These changes indicate a decrease of the inter-particle distance as well as an increase of the local ordering (compare with Figure1d and 1e). The initial RDF of the system in the absence of blue light ($t$ = 0 s, green curve in Figure 2k) exhibits only one single peak, followed by a featureless signal, which is typical for amorphous materials. The appearance of discrete peaks (indicated with orange arrows) in the RDF taken for the particle distribution after illumination with blue light for $t$ = 0.7 s and $t$ = 1.4 s (Figure 2k) is in line with denser media and enhanced order. However, the matrix is still in the amorphous state even in the region of highest particle density. With the available strength of the catalytic activity of the defect and having open boundary conditions, we do not achieve the critical concentration of passive beads to realize the transition into the crystalline state.

**Higher density of active defects:** To explore the structural phase transition from the amorphous to crystalline state, we experiment with systems possessing different number of active defects. The assumption is that the radial compression waves from each defect could lead to a higher local density of passive beads, which will be formed between the defects. To set up this experiment, we bring active Janus particles on a microfluidic chip, consisting of multiple microwells and monitor transients of the passive colloidal matrix for the case when the system has 2 and 6 active defects (Figure 3). In the following experiments, we keep the concentration of passive beads constant by placing them in a microwell. The microwell is taken sufficiently large (diameter of about 300 µm and height of 15 µm) that it does not affect the processes of interest but only assures that the average particle density remains constant. For this purpose, we immobilize several active Janus defects at a glass substrate inside the microwell, to reach less than 10 Janus particles in each well. With this density, we achieve an average distance between Janus particles of about the critical compression distance $L_c$ ~ 22 µm. In this way, we assure that the passive matrix in between neighboring Janus beads will be compressed by the active defects. Finally, passive silica beads are added with the nominal concentration of 0.176 $\mu m^{-2}$. The microwell is illuminated with a blue light and the spatial distribution of passive beads is analyzed as a function of the illumination time.

Panels a to c in Figure 3 show micrographs for the case when there are only 2 Janus particles in the field of view at a distance ~ 55 µm, surrounded by the silica passive beads. The distance between active defects is slightly larger than twice the critical propagation distance of the particle density wave, which is about 45 µm (~ 2 $L_c$). Before the sample is illuminated with blue light ($t$ = 0 s, Figure 3a), passive beads are homogenously distributed and form an amorphous matrix. Once the blue light is turned on, the exclusion between the two active defects and passive beads leads to a compactification of the passive matrix (Figure 3b and 3c). The highest density of passive beads achieved in the area between the defects is 0.195 $\mu m^{-2}$ (Figure 4), which is larger than that for the case of a single active defect but still sub-critical to initiate the crystallization process.

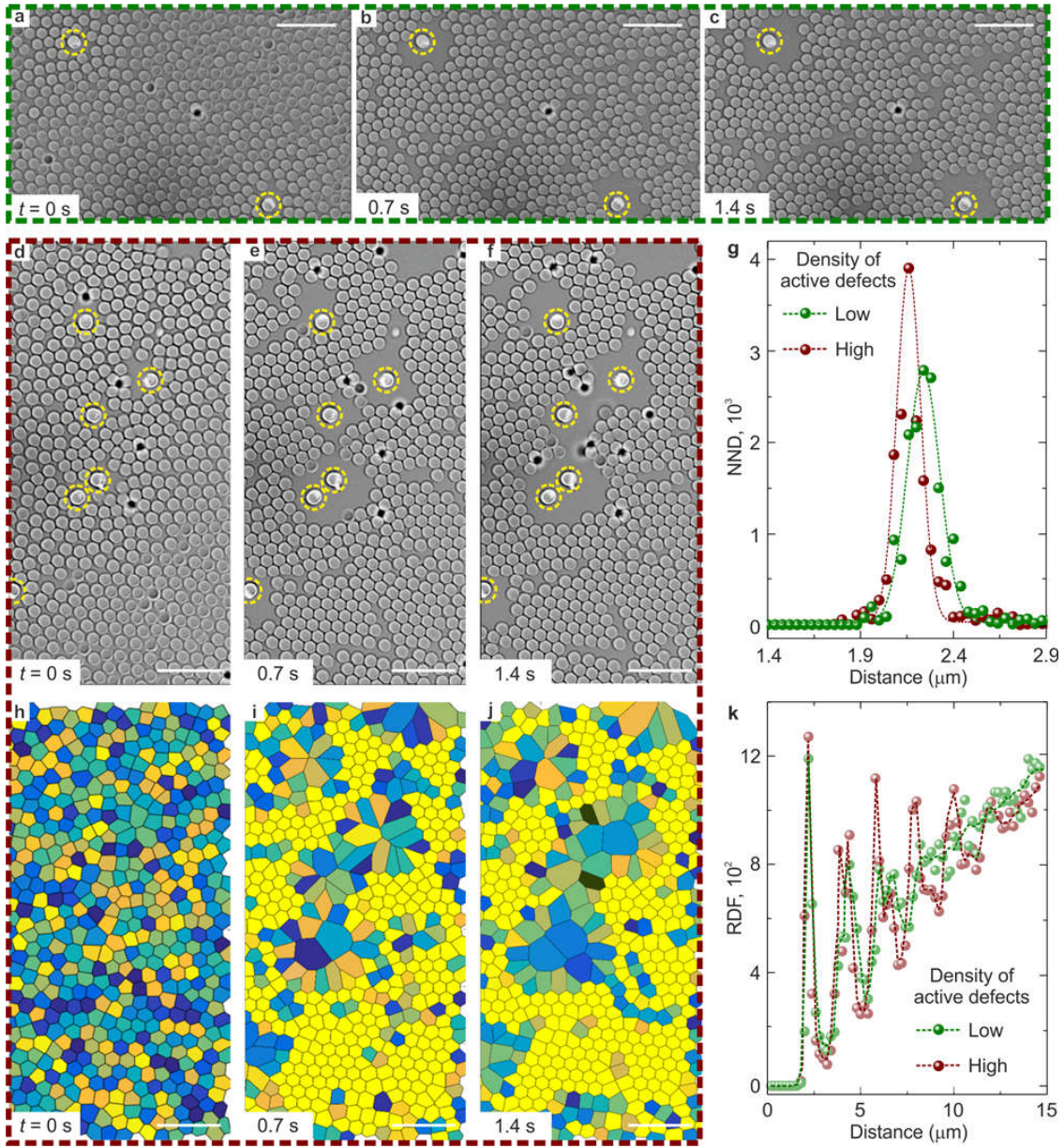

**Figure 3**. Formation of a crystalline state in an amorphous passive matrix due to different number of active defects (experiment). (a-c) A sequence of optical microscopy images showing the interaction between 2 immobile Janus particles (indicated with yellow dashed circles) and a dense passive matrix. Scale bar, 10 μm. The density of passive beads increases in the region between the active defects but remains sub-critical for crystallization. (d-f) A sequence of optical microscopy images showing the impact of 6 active defects (indicated with yellow dashed circles) on the initially amorphous matrix of passive silica beads. Scale bar, 10 μm. (h-j) Voronoi diagrams illustrating the emergence of the crystalline order under blue light illumination, where each cell corresponds to a particle. Voronoi diagrams for the data shown in (a-c) are presented in Figure S6. Voronoi diagrams (h-j) correspond to the images shown in (d-f). Scale bar, 10 μm. (g) The NND curves and (k) the RDF calculated after 0.7 s for the case of low density (2 particles) and high density (6 particles) of active defects. A higher density of active defects facilitates the emergence of long-range order in the system and local crystallization.

The situation changes drastically when the concentration of active defects is increased to six and the inter-defects distance becomes smaller than $L_c$. Once illuminating the sample for 0.7 s with blue light, we observe the crystallization of passive beads in the area between active Janus particles (Figure 3, d to f). This statement about the onset of a phase transition is supported by the following quantitative analysis: Figure 3h-j, shows Voronoi diagrams[24] that visualize the transient regime from the amorphous to crystalline state. The hexagonal yellow-colored cells correspond to the regions where colloidal beads form a close-packed hexagonal lattice driven by the internal compression. Other colors indicate either cells with different number of edges or non-ideal hexagons (*e.g.*, when a particle is at the edge of the assembly). Further, we analyze the distribution of passive beads at $t = 0.7$ s and compare the NND (Figure 3g) and RDF (Figure 3k). The RDF shows multiple sharp peaks and the NND narrows down similar to the case sketched in Figure 1b and 1e, due to the higher degree of long-range order in the system. The highest density of passive beads in this experiment is 0.216 $\mu m^{-2}$. This concentration is just above the critical one to achieve crystallization.

The experimentally observed *compression* of the colloidal matrix is associated with the local decrease of the interparticle distances around the defect (Figure 2 and Figure 3). The pressure on the passive matrix decays with the distance from the active defect.[9] For the case of photocatalytic Janus particles with a diameter of 2 μm, the critical propagation distance of the particle density wave from the defect is determined to be around $L_c = 22$ μm (Figure 2i). This is a consequence of the specific activity of the defect, which is quantified[22b] to be $\gamma = 3.5$ $\mu m^2$/s. The phenomenological parameter γ (see Eq. (4) in the Simulations subsection below) is a cumulative "strength of the flow",[22b] which includes the flow of the fluid, the gradient of the concentration profile of ions, and other products of the chemical reaction from the surface of the AgCl particles[25] illuminated by blue light. If a bead is initially located at a distance ≥ $L_c$ from the active defect, it will stay unaffected by the local photochemical reaction and will exhibit a conventional Brownian motion. Still, within a ring-shaped area around the active defect with an outer radius $L_c$, the density of passive beads is enhanced compared to the initial value. If the density of passive beads will overcome a certain threshold, the system will crystallize. By analyzing the RDF and NND dependencies, we can monitor the transition from the amorphous to crystalline states in the system. Experimentally, we clearly see that by increasing the number of active defects, the system reveals a tendency towards global crystallization (Figure 3d-j).

To validate the possibility of global crystallization of the entire passive matrix due to the internal drive, we performed molecular dynamics simulations of a colloidal system consisting of an array of active defects surrounded by high-density passive beads (Figure 4). We model the situation when the average density of passive beads remains constant (the initial density is 0.157 $\mu m^{-2}$). Active defect is represented by a Janus particle characterized by $\gamma = 3.5$ $\mu m^2$/s. Each simulation cell has size 33 × 33 $\mu m^2$ and contains one active defect. We employ periodic boundary conditions, thus allowing particles to freely travel between the neighboring cells. Employing periodic boundary conditions means modeling an infinite array of cells each containing one (or more) active defect. To illustrate the formation of crystals between the defects, we show in the plots an area accommodating nine defects: a simulation cell and eight surrounding images. Figure 4a shows a snapshot of the particle distribution at the initial time $t = 0$ s, before the defects become catalytically active. Passive beads are randomly distributed around Janus defects, similar to the case shown in Figure 2a, with an average density of 0.157 $\mu m^{-2}$. The corresponding analysis of the RDF reveals only first characteristic peak typical for disordered systems (Figure 4d). Furthermore, the NND (Figure 4e) is very broad, indicating no long-range order in the system. Once the defects become catalytically active, they start ejecting ions that repel passive beads in their vicinity. At the initial stages of the exclusion process ($t = 0.1$ s) passive beads start to be compressed, leading to an increase of the density to the maximum value of 0.171 $\mu m^{-2}$. The corresponding RDF is minimally affected compared to the initial situation, while the NND is sharpened, featuring a suppressed long tail. With time, the compression wave propagates in the dense environment and already at $t = 1$ s the RDF reveals a clear 3-peak structure accompanied by the appearance of a sharp peak in the NND (Figure 4b, f and g). This is indicative of the onset of long-range order in the system. The density of passive beads is increased to the maximum value of 0.181 $\mu m^{-2}$, which is still sub-critical, and the system shows only a local crystalline order. A crystallization of the initially amorphous matrix is achieved after about 10 s at a constant drive due to the chemical reaction (the density of passive beads reaches a value as high as 0.205 $\mu m^{-2}$ in the crystalline state) and is well-resolved in the multi-peak RDF profile and sharpened NND function (Figure 4c, h and i). The maximum of the NND is slightly shifted towards shorter distances as the local inter-particle distance decreases and the long-distance tail disappears. With this, the system reveals a global amorphous-to-crystalline phase transition. Note that the resulting solid is porous because the total density of particles in the system remains constant.

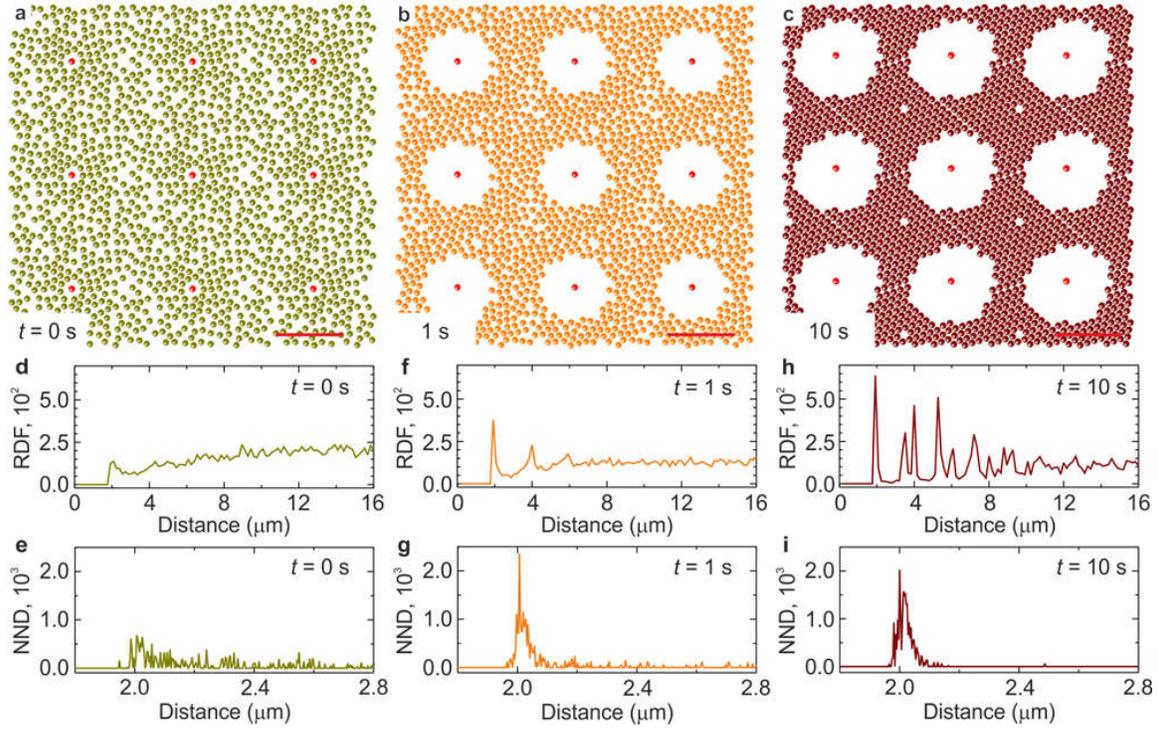

**Figure 4**. Emergence of a global crystalline state from an amorphous matrix due to internal drive (simulation). (a-c) Time evolution of the distributions of passive beads in the presence of active defects (shown for one simulation cell and eight images). Scale bar, 20 μm. (a) Snapshot of the particle distributions at the initial time $t = 0$ s (amorphous state). The corresponding RDF only shows a smooth first peak and then is featureless (d), and the NND shows no clear peak (e). (b) The particle distribution at $t = 1$ s: The corresponding RDF (f) and NND (g) show the onset of long-range order. (c) Formation of a crystalline solid at $t = 10$ s: The corresponding RDF (h) shows many sharp peaks, and the NND (i) is represented by a narrow peak.

The observed active defect-induced structural phase transition is distinct from the phase transitions studied in amorphous and crystalline materials. For example, in elastic colloidal crystals, the propagation of localized mechanical pulses in aqueous monolayers of micron-sized particles of controlled microstructures leads to the melting of a crystal.[9, 26] This is in line with the second law of thermodynamics, where defects destroy the local crystalline order and lead to crystal melting. This situation is exactly the opposite to the scenario reported in our work, where the initially amorphous (or liquid) matrix acquires a crystalline order thanks to the photocatalytic activity of Janus defects, which produce an internal driving force for crystallization.

There is only one key parameter governing the crystallization process, namely the defect activity $\gamma$ (see Eq. (4) in Simulations and Ref. [22b]). It determines the critical propagation distance of the particle density wave and, hence, defines the distance at which active defects should be located to assure crystallization. For our system of silica beads with photochemically driven Ag/AgCl Janus particles, the distance between active defects should be about 45 μm. The parameter $\gamma$ is specific to the system of choice. For instance, its value can be tailored by the chosen photocatalytic material, intensity of illumination, or by the size of the object. For the latter, as follows from Figure 3d-j, it is insightful to position several active defects close to each other to effectively enhance $\gamma$. Using larger agglomerates of Ag/AgCl Janus particles allows to increase $\gamma$ by about an order of magnitude, compared to the case of a single Janus particle.[22b]

## 3. Discussion

The physical interpretation of the observed transition from amorphous/liquid to solid state induced by active defects depends on the choice of the continuously varying parameter governing the behavior of the system. Below, we present two possible scenarios. However, we should keep in mind that the system at hand is out of equilibrium, and one cannot directly apply thermodynamic fields from equilibrium thermodynamics to define the transition. Moreover, the observed crystallization with increasing activity of defects is rather a process opposite to what the equilibrium thermodynamics implies. Indeed, this process is accompanied by an increase in the order, and thus decrease in the entropy, at increasing defect activity, which is opposite to the melting transition[2, 9]. This situation can only be realized at conditions far from the thermodynamical equilibrium as in our case the energy of the ordered crystalline state is higher than that of the initially disordered state. To remain in the crystalline state, the system requires a constant pump of energy, which is supplied in our case from the catalytic reaction. Such non-equilibrium states can be defined as induced states and they are known in physics, in particular, in quantum physics. For example, by driving a system of polaritons in semiconductors by laser radiation, the system can display an induced Bose-Einstein condensate (BEC)[27] which, in contrast to the usual BEC, is not the lowest energy state of the system.

Taking the non-equilibrium nature of the resulting solid, one can map the observed phenomenon on the corresponding processes in equilibrium thermodynamics. One way to do so would be to consider the transient pressure exerted on passive beads by active defects and described by the phenomenological parameter $\gamma$ introduced in our simulations. In terms of the pressure defined in this way, the observed transition looks as a pressure-induced direct transition from amorphous/liquid to solid at increasing pressure. This is a very illustrative interpretation and it provides a direct link to the blast-induced solidification mentioned above.

However, this interpretation should be taken with care due to the non-equilibrium nature of the resulting solid state. The transient pressure results from the flow of ions from the surface of active defects. It is not a steady-state flow, as it changes in time. Also, due to the fast re-arrangement of the particle distributions in the vicinity of the active defect, the flow pattern is complex and rapidly changing. Note also that the ion flow leaks through the gaps between spherical passive beads and rapidly dissipates outside the circle with the radius $L_c$. The ions emitted from the surface of a Janus particle can induce their self-propelled motion (in case, when they are mobile). The flows of ions, which compress passive beads and ultimately squeeze them into a solid, are a non-equilibrium system. The thermodynamics of gases and, in particular, the Equation of State (EOS) is not applicable in general case to active systems (see Ref. [28] and references therein). However, considering it as an approach to map our observation on the equilibrium thermodynamics, the interpretation in terms of transient pressure seems to be a good approximate approach for the understanding of the observed solidification.

An alternative approach implies that, in addition to thermal fluctuations responsible for the thermal Brownian motion, there are fluctuations induced by active defects, and their amount is directly controlled in the experiment by the intensity of light. Note that these additional fluctuations are responsible for substantially larger values of the MSD of passive beads in the vicinity of a Janus particle (the exclusion effect[22b]). The motion of passive beads is therefore due to two sources of fluctuations: thermal noise exerted on the bead by the environment (thermal Brownian motion) and the kick exerted by the active defect (immobilized Janus particle). Therefore, exactly in the same way as for a self-propelled particle[22b, 25], the total effective diffusion of a passive bead near the surface of an active defect (inside a circle of radius $L_c$) can be expressed:

$$D_{pb,eff} = D_{pb,0} + v^2 \tau_{r,pb}/4, \qquad (1)$$

where $D_{pb,0}$ is a Brownian diffusion constant of a bead, $\tau_{r,pb}$ is its inverse rotational diffusion coefficient, and $v$ is the additional velocity of the bead due to the push from the active defect. The "effective temperature" $T_{eff}$, corresponding to the total effective diffusion $D_{pb,eff}$, can be found from:

$$D_{pb,eff} = (k_B T_{eff})/(6\pi\eta r_{pb}), \qquad (2)$$

here $\eta$ is the fluid viscosity. The effective temperature $T_{eff}$ is a measure of the sum of fluctuations exerted on a passive bead from (i) the active defect and (ii) from other environment.

If we initially set the effective temperature $T_{eff}$ very low, the system will be in solid state (not necessarily crystalline), i.e., passive beads will only fluctuate near their initial positions but will not move to other cells. With increasing the effective temperature $T_{eff}$ (either by increasing temperature of the system or the intensity of light), we turn the system to liquid, so the beads execute Brownian motion. With further increase of $T_{eff}$ (by increasing the light intensity) the beads diffuse faster, and the disorder in the system further increases. However, when the light intensity increases even further, such that the radii of exclusion of neighboring Janus particles overlap, the system turns to a crystal. Therefore, the system undergoes first a transition from solid to liquid and then from liquid to solid when continuously increasing the thermodynamic field $T_{eff}$. Therefore, this scenario describes a *re-entrant* solidification with respect to $T_{eff}$. Note that in our experiment we do not observe the initial solid state, which can be achieved at low $T_{eff}$.

Alternatively, if we introduce a weak repulsion between the beads, then the same reentrant behavior can be achieved even without lowering the system temperature. In this case, we can fix the temperature of the environment and only increase the intensity of light to observe the reentrant transition. Indeed, the repulsion between the beads would order them in a hexagonal crystal lattice when the light is off. With increasing intensity of the light, the crystal will first melt and then undergo the transition to solid again, when the intensity of the light is sufficiently high. We note that for a system of weekly repulsive beads, a reentrant solidification should be observed for a pressure-driven and fluctuation-driven scenarios. Indeed, at zero activity of active defects (light is off) in both cases passive beads will arrange themselves in a hexagonal lattice.

It is also worth noting that the active defect-induced transition revealed in our work, is conceptually similar to the "freezing-by-heating" transition[3, 5] observed in a system of driven colloids. The "freezing-by-heating" transition requires three ingredients.[3, 5] external driving keeping the system out of equilibrium, repulsive interaction among the species, and confinement. In our case, the role of external driving is played by the chemical activity of active defects immersed in a passive matrix. All the particles in the system, active and passive, possess a core-to-core repulsive interaction in addition to the long-range repulsion exerted on the particles from the activity of the defects. The confinement is effectively imposed by a constant density of passive particles in the system, which is modeled by periodic boundary conditions in the simulations. Therefore, we can call the observed solidification as an active-matter realization of the unusual "freezing-by-heating" transition.[3, 5] Same as the "freezing-by-heating", the transition observed in this work, has a potential for many interesting applications, for example, for studying pedestrian motion and traffic patterns including jamming and panic escape. Indeed, the compressed solid state can be considered as a fully jammed state of species trying to escape from an area affected by active defects (modeling, e.g., a panic escape of pedestrians from fire or other dangerous impacts, or escape of birds from fireworks areas in cities or animals from forest fires). By allowing a leakage of particles through small openings, we can model the possibility of the escape through gates. But to make such demonstrations more efficient, both, active and passive particles should be mobile.

## 4. Conclusions

We demonstrated a novel solidification transition in an initially liquid or amorphous system doped by active defects, due to internal drive stemming from active defects. This scenario is validated experimentally and in simulations with a soft matter model system, consisting of photocatalytically active yet immobile Janus particles acting as defects in a dynamically reconfigurable matrix of passive colloidal beads. The two sub-systems interact repulsively due to the efficient photochemical production of ions generating a local chemical gradient around Janus particles, resulting in an exclusion behavior of the dense matrix of passive beads. We show that the exclusion effect can result in compression of the passive colloidal matrix. In stark contrast to the behavior of systems in equilibrium, the initially amorphous colloidal matrix undergoes an amorphous-to-crystalline phase transition when increasing the density and/or activity of active defects. This unusual induced crystallization is specific to systems which are driven out of equilibrium, and can be achieved if the density of active defects overcomes a certain threshold related to the specific activity of the defect. We discussed possible interpretations of the observed amorphous-to-solid transition: in terms of transient pressure exerted on passive beads from active defects and in terms of effective temperature as a measure of thermal fluctuations and fluctuations due to the catalytic activity of defects. The observed induced solidification can be also considered as an active soft-matter realization of the so-called "freezing-by-heating" transition earlier observed in an externally driven system. Indeed, in both cases fluctuations (in our case, induced by active defects) result not in conventional melting but in an inverse solidification, or "freezing" of the initially disordered, i.e., liquid or amorphous, state.

The prediction of the active defect-induced structural phase transition due to internal drive is generic and is not limited to the 2D systems studied here. It can be extended to address non-equilibrium crystallization processes in 3D systems, where the crystalline phase will be formed between active defects assembled in a 3D array. The lattice of active defects does not need to be regular. It is sufficient that neighboring active defects are positioned within the critical propagation distance of the particle density wave, which is determined by the activity of the defect. Our proposal can be readily combined with the appealing concepts involving mobile active Janus particles, which can propagate along the boundaries between crystallites and assist the formation of large area defect-less crystals.[29]

The observed amorphous-to-crystalline phase transition provides a novel insight into the collective effects in mixed colloidal systems of active and passive species. It offers versatile possibilities to address the processes of solidification in various systems brought out of equilibrium, including the formation of biomolecular condensates or biomineralization, transitions from amorphous to polycrystalline state in condensed matter, or the synthesis of materials under extreme conditions. We demonstrate that crystallization via internal drive due to the presence of active defects is accompanied by porosity of the resulting material. In this respect, the proposed crystallization mechanism allows to tailor mechanical properties of new functional materials by maintaining their electronic properties.

## 5. Experimental Section

*Materials and Instruments:* The used chemicals including PVP (Mw = 55 000), Iron(III) chloride hexahydrate ($FeCl_3 \cdot 6H_2O$), and the polystyrene and $SiO_2$ microparticles (diameter of 2 μm) are from Sigma-Aldrich. The fluorescence lamp (HBO 103) is from Carl Zeiss Microscope (Mercury lamp, blue light, 450-470 nm). The light power detector and its monitor are from Gentec-eo.

*Janus particles preparation:* Janus particles are fabricated following the procedure by Wang et al.[22] In brief, monolayers of polystyrene (PS) spheres with a diameter of 2 μm are prepared by casting a drop of colloidal suspension onto thin glass substrates. Then, a silver (Ag) layer of about 60 nm is deposited onto the surface of PS particle monolayers by thermal evaporation at a base pressure $7 \times 10^{-5}$ mbar. Afterwards, Janus particles are detached from the substrate using an ultrasonication process and resuspended in deionized water. Particles are further dispersed into a polyvinylpyrrolidone (PVP) solution (300 mM) for the synthesis of Ag/AgCl layers. The synthesis process of Ag/AgCl is conducted in a dark environment in a solution with an excess concentration of $FeCl_3$ (20 mM) during 60 min. Finally, colloids with the synthesized Ag/AgCl hemispheres are washed in deionized water using centrifugation process, and then dispersed in deionized water for further experiments.

*Blue light illumination*: The blue light illumination (wavelength of 450-470 nm; power density of 137 μW $mm^{-2}$) is realized using an external fluorescence lamp with a Zeiss filter set.

*Substrate cleaning:* The glass slides as substrates are pretreated by immersion in a hot $H_2SO_4/H_2O_2$ (7 : 3) solution for 30 min, followed by washing with a copious amount of deionized water.

*Preparation of microwells:* Microwell structures (15 μm high and diameter of 300 μm) are fabricated using a negative photoresist (SU-8 2010, Microchemicals, Westborough, USA) on a glass substrate. After substrate cleaning, an adhesion promoter (TI Prime, Micro chemicals GmbH, Germany) is coated at 4000 rpm for 50 s. Then, 15 μm thick SU-8 is coated, prebaked at 95°C for 3 min, and exposed to UV light (Karl Süss MJB4, Garching, Germany). After post-baking at 95°C for 6 min, the structure is developed (mr600, Microresist Technology GmbH, Germany), cleaned, and finally baked at 120°C for 30 min.

*Radial distribution function:* The algorithm involves determining how many beads are within the area defined by two circles with radii $r$ and $r + dr$ away from the active Janus particle, followed by calculating the distance between all particle pairs and binning them into a histogram.

*Calculation of the particle density:* The image processing software Fiji (http://fji.sc/) is used for counting the number of particles, $N$, and calculating the area, $A$, for the selected image. The density of particles is calculated as $N/A$.

*Video analysis:* TrackMate, from the image processing software Fiji (http://fji.sc/), is used for particle tracking to obtain the trajectory data.[30]

*Simulations:* The behavior of the system consisting of active Janus particles and passive beads is simulated by numerically integrating the overdamped Langevin equations:[10d, 22, 31]

$$\dot{x}_i = v_0 \cos\theta_i + \xi_{i0,x}(t) + \sum_{ij}^{N} f_{ij,x},$$
$$\dot{y}_i = v_0 \sin\theta_i + \xi_{i0,y}(t) + \sum_{ij}^{N} f_{ij,y}, \quad (3)$$
$$\dot{\theta}_i = \xi_{i\theta}(t),$$

for $i, j$ running from 1 to the total number $N$ of particles, active and passive; $v_0$ is self-velocity of Janus particles which is set to zero for immobilized particles. Here, $\xi_{i0}(t) = (\xi_{i0,x}(t), \xi_{i0,y}(t))$ is a 2D thermal Gaussian noise with correlation functions $\langle \xi_{0,\alpha}(t) \rangle = 0$, $\langle \xi_{0,\alpha}(t)\xi_{0,\beta}(t) \rangle = 2D_T \delta_{\alpha\beta}\delta(t)$, where $\alpha, \beta = x, y$ and $D_T$ is the translational diffusion constant of a passive particle at fixed temperature; $\xi_\theta(t)$ is an independent 1D Gaussian noise with correlation functions $\langle \xi_\theta(t) \rangle = 0$ and $\langle \xi_\theta(t)\xi_\theta(0) \rangle = 2D_R \delta(t)$ that models the fluctuations of the propulsion angle $\theta$. The diffusion coefficients $D_T$ and $D_R$ can be directly calculated or extracted from experimentally measured trajectories and MSD, by fitting to theoretical MSD.[22b] Thus, for a particle with diameter of 2 μm diffusing in water at room temperature, $D_T \approx 0.22$ μm$^2$ s$^{-1}$ and $D_R \approx 0.16$ rad$^2$ s$^{-1}$.

The term, $\sum_{ij}^{N} f_{ij}$, represents, in a compact form, the sum of all inter-particle interaction forces in the system including[22b]: (i) elastic soft-core repulsive interactions between active particles, between passive beads, and between active and passive particles; and (ii) the effective repulsive interaction between Janus particles and passive beads due to the radial flow of products of chemical reaction from the surface of Ag/AgCl Janus particles illuminated by blue light.

The latter contribution (ii) is modeled by a finite-range field of radial forces, decreasing in amplitude as $1/r$ from the center of a Janus particle:

$$f_{ij}^{flow} = \begin{cases} \dfrac{\gamma}{|\vec{r}_i - \vec{r}_j|}, & \text{if } |\vec{r}_i - \vec{r}_j| > R_i + R_j, \\ 0, & \text{if } |\vec{r}_i - \vec{r}_j| \gg R_i + R_j, \end{cases} \quad (4)$$

where $\gamma$ is the cumulative "strength of the flow" parameter, which includes the flow of the fluid, the gradient of the concentration profile of ions and other products of the chemical reaction from the surface of the AgCl particles illuminated by blue light. From the calibration of the simulated trajectories and the MSD of beads to the experimentally measured MSD we found[22b] that for single Janus particles $\gamma \approx (3.2 - 3.9)$ μm$^2$ s$^{-1}$.

**Acknowledgements**

The authors thank B. Kruppke (TU Dresden) for support with the SEM measurements, B. Ibarlucea (TU Dresden) and A. Caspari (IPF) for Zeta potential measurements. This work was supported in part via the German Research Foundation (DFG) via DFG grants BA 4986/7-1, MA 5144/9-1, MA 5144/13-1, MA 5144/14-1. T.H acknowledges the China Scholarship Council (CSC) for financial support. V.R.M. and F.N. acknowledge support by the Research Foundation-Flanders (FWO-Vl) and Japan Society for the Promotion of Science (JSPS) (JSPS-FWO Grant No. VS.059.18N). F.N. is supported in part by the MURI Center for Dynamic Magneto-Optics via the Air Force Office of Scientific Research (AFOSR) (FA9550-14-1-0040), Army Research Office (ARO) (Grant No. Grant No. W911NF-18-1-0358), Japan Science and Technology Agency (JST) (via the Q-LEAP program, and the CREST Grant No. JPMJCR1676), Japan Society for the Promotion of Science (JSPS) (JSPS-RFBR Grant No. 17-52-50023, and JSPS-FWO Grant No. VS.059.18N), the RIKEN-AIST Challenge Research Fund, the Foundational Questions Institute (FQXi), and the NTT PHI Laboratory.